\documentclass[amsmath,amssymb,aps,prx,reprint,superscriptaddress,longbibliography,floatfix]{revtex4-2}
\usepackage{amsmath}
\usepackage[table]{xcolor}
\usepackage{amsfonts,amssymb}
\usepackage{graphicx}
\usepackage{float}
\usepackage{hyperref}
\usepackage{academicons}
\usepackage{multirow}
\usepackage[version=4]{mhchem}
\usepackage{makecell}
\usepackage{pgf}
\usepackage{booktabs}
\usepackage[english]{babel}
\usepackage[normalem]{ulem}
\usepackage{natbib}
\usepackage[ruled,linesnumbered]{algorithm2e}

\newcommand{\method}[1]{\noindent\textbf{#1.}}

\newcommand{\VJP}{\operatorname{VJP}}
\newcommand{\Zpair}{\mathbf Z^{*}_{\text{pair}}}
\newcommand{\tq}{q^{\text{raw}}}
\newcommand{\qraw}{\mathbf q^{\text{raw}}}

\SetAlgoCaptionSeparator{\unskip}
\SetAlgorithmName{\unskip}{algorithm}{List of algorithms}

\SetAlgoCaptionLayout{algocapleft}



\colorlet{cmin}{green!10!white}
\colorlet{cmed}{yellow!20!white}
\colorlet{cmax}{red!20!white}
\newcommand{\ColorCellRange}[3]{%
  \begingroup
  \pgfmathsetmacro{\norm}{max(min((#3-#1)/(#2-#1),1),0)}%
  \pgfmathtruncatemacro{\pctint}{round(\norm*100)}%
  \ifnum\pctint<50
    \pgfmathtruncatemacro{\mix}{round(100 - 2*\pctint)}%
    \edef\temp{\noexpand\cellcolor{cmin!\mix!cmed}}%
  \else
    \pgfmathtruncatemacro{\mix}{round(200 - 2*\pctint)}%
    \edef\temp{\noexpand\cellcolor{cmed!\mix!cmax}}%
  \fi
  \temp #3%
  \endgroup
}
\newcommand{\CE}[1]{\ColorCellRange{0}{4.1}{#1}}
\newcommand{\CF}[1]{\ColorCellRange{0}{96}{#1}}
\newcommand{\CZ}[1]{\ColorCellRange{0}{0.48}{#1}}

\begin{document}

\title{Scaling charge-aware machine learning molecular dynamics through \\memory-bounded GPU streaming}

\author{Xiaoyun Yu}
\affiliation{Department of Materials Science and Engineering, National University of Singapore, Singapore}

\author{Yicheng Chen}
\affiliation{Department of Materials Science and Engineering, National University of Singapore, Singapore}

\author{Peichen Zhong}
\email[]{zhongpc@nus.edu.sg}
\affiliation{Department of Materials Science and Engineering, National University of Singapore, Singapore}

\date{\today}

\begin{abstract}
Charge-aware machine learning interatomic potentials (MLIPs) capture long-range electrostatics and electrical response, but retaining neural activations across the global electrostatic solve creates an activation memory wall that limits simulation scale. Here we introduce ERStream, a GPU engine for learned-charge equivariant MLIPs, applied to one- and two-layer MACE-ER and DPA4C-ER. ERStream discards forward activations before the global solve and rematerializes them per tile afterward, confining neural memory to one tile without altering the trained potential. Evaluating long-range forces and Born-effective-charge response at every step, one-layer MACE-ER accommodates 300 million water atoms on one NVIDIA B200 and 81 million on one RTX~5090, while eight GPUs achieve a 7.8-fold speed-up for a 2.1-million-atom solvated protein. Trained solely on energies and forces, these models reproduce SrTiO$_3$ phonon splitting and water-cluster infrared spectra. ERStream also runs explicit-solvent protein dynamics, a 10-million-atom isolated water droplet, and solid--liquid wetting. Removing this activation memory wall brings device-scale simulations of collective electrostatic, interfacial, and field-driven phenomena within reach of routine atomistic modeling.
\end{abstract}

\maketitle
\raggedbottom
\clubpenalty=10000
\widowpenalty=10000
\displaywidowpenalty=10000
\setcounter{topnumber}{3}
\setcounter{totalnumber}{4}
\setcounter{dbltopnumber}{2}
\renewcommand{\topfraction}{0.95}
\renewcommand{\dbltopfraction}{0.95}
\renewcommand{\textfraction}{0.05}
\renewcommand{\floatpagefraction}{0.65}
\renewcommand{\dblfloatpagefraction}{0.65}
\makeatletter
\setlength{\@fptop}{0pt}
\setlength{\@fpsep}{14pt plus 2pt minus 2pt}
\setlength{\@fpbot}{0pt plus 1fil}
\setlength{\@dblfptop}{0pt}
\setlength{\@dblfpsep}{14pt plus 2pt minus 2pt}
\setlength{\@dblfpbot}{0pt plus 1fil}
\makeatother

\section{Introduction}

Long-range electrostatic interactions couple atomic motion to collective thermal and electrical response.
Macroscopic polarization~\cite{falletta2025}, ion transport~\cite{gomez2024}, and dielectric response to external fields~\cite{joll2024, montenegro2021, wang2025water} emerge from this coupling through correlations extending beyond local atomic environments~\cite{gao2022}.
Simulating these phenomena requires \textit{ab initio} accuracy, explicit long-range electrostatics and electrical response, and device-scale system sizes.
Machine learning interatomic potentials (MLIPs) approximate density functional theory (DFT) potential energy surfaces at linear cost in system size by decomposing energies into local environments with finite cutoffs or bounded message-passing depths~\cite{behler2007, bartok2010, deepmd2018, mace2022, allegro2023sc}.
This locality has enabled Deep Potential molecular dynamics (MD) simulations of 100 million atoms~\cite{jia2020}, but it omits electrostatic coupling and electrical response beyond the finite receptive field~\cite{grisafi2019, anstine2023}.

To capture long-range electrostatics, recent MLIPs incorporate charge equilibration~\cite{ko2021}, learned Wannier centers~\cite{dplr2022}, or latent charges inferred directly from energies and forces~\cite{les2024, les2025}.
Latent Ewald Summation links these charges to macroscopic polarization and Born effective charges (BECs)~\cite{zhong2025}, whereas differentiable field-dependent potentials offer an alternative route to electrical response~\cite{falletta2025}.
We previously augmented the strictly local DPA4C descriptor~\cite{dpa4c2026} with learned-charge electrostatics, factoring long-range forces and BECs into one system-wide electrostatic calculation (hereafter, the global solve) and local charge derivatives~\cite{dpa4clr2026}.
Yet scaling these charge-aware MLIPs to device-sized systems remains an open challenge.

For short-range potentials, recent GPU engines reach device scale along two routes: multi-GPU parallelism and single-GPU streaming.
Spatial domain decomposition distributes Deep Potential, Allegro, and DPA4C across GPUs~\cite{jia2020, allegro2023sc, dpa4c2026}, whereas SevenNet, DistMLIP, and Tessera parallelize multi-hop graph neural networks~\cite{sevennet2024, dispmlip, tessera2026}.
On a single GPU, Symmetrix-XL evaluates 11 million atoms with a pretrained short-range MACE model in 80~GB of memory by streaming sequential subsets of atoms (tiles)~\cite{symmetrixxl}.
Operator libraries such as cuEquivariance and NVIDIA ALCHEMI Toolkit-Ops further accelerate equivariant tensor products and atomistic kernels~\cite{cuequivariance, nvalchemi050}.
Both scaling routes rely on locality, allowing each domain or tile to be evaluated independently from a finite surrounding neighborhood (halo).

Charge-aware MLIPs break this locality at the global solve, where all learned charges enter one system-wide electrostatic calculation that returns electrostatic derivatives (charge adjoints) to every atom.
Because no domain or tile can complete backpropagation for forces until the solve finishes, 
retaining the forward neural activations of all atoms across the solve creates a GPU memory wall~\cite{gholami2024}.
Computing BECs and electrical response further requires the position Jacobian of the learned charges, and equivariant message passing compounds both memory and derivative costs through high-dimensional tensor features and multi-hop dependencies.
To date, charge-aware MD with BECs has reached million-atom systems on consumer GPUs with the neuroevolution potential qNEP~\cite{qnep2026} and 5.2 million atoms on four GPUs with our strictly local equivariant model~\cite{dpa4clr2026}.
Matching short-range engine scale without altering the trained potential requires bounding per-GPU memory, distributing execution across GPUs, and computing response derivatives across the global solve.

\begin{figure*}[htb]
\centering
\includegraphics[width=\textwidth]{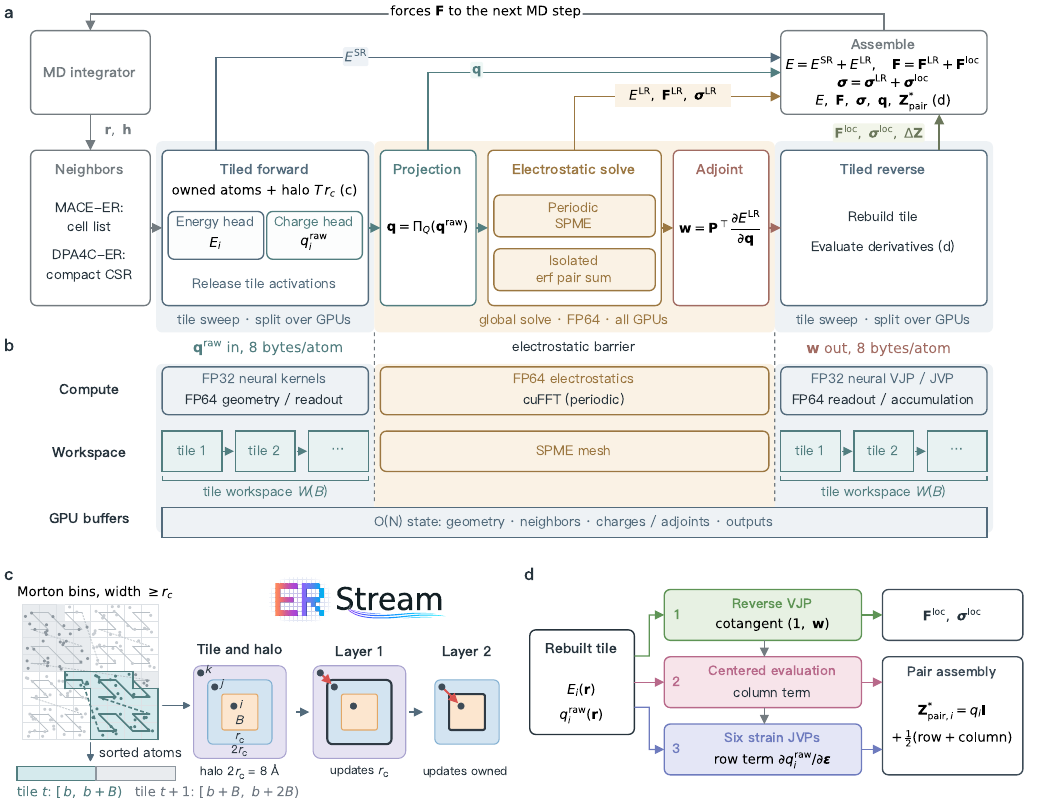}
\caption{\textbf{ERStream partitions each MD step at the electrostatic barrier.} \textbf{a}, Step data flow. The tiled forward sweep writes site energies $ E_i$ and raw charges $\tq_i$ and discards activations. After neutrality projection $\Pi_Q$, the global solve (periodic SPME or isolated erf sum) returns adjoint $\mathbf w$. The tiled reverse sweep rebuilds each tile for local forces $\mathbf F^{\text{loc}}$, stress $\boldsymbol\sigma^{\text{loc}}$ and charge-response term $\Delta\mathbf Z$ of $\Zpair$. 
\textbf{b}, Compute, workspace and GPU buffers aligned with \textbf{a}. Only $\qraw$ and $\mathbf w$ cross the global stage (8~bytes per atom each). Tiles reuse one workspace of peak size $W(B)$, leaving only $O(N)$ buffers resident. 
\textbf{c}, ERStream tiling and halo, shown for MACE-ER-T2. Morton-sorted tiles of $B$ owned atoms are rebuilt with inner ($r_c$, blue) and outer ($2r_c$, lavender) shells, with updated regions outlined in bold and messages in red. 
\textbf{d}, Sequential passes on one rebuilt MACE-ER-T2 tile: reverse VJP for $\mathbf F^{\text{loc}}$ and $\boldsymbol\sigma^{\text{loc}}$, centered evaluation for the column term (Eq.~\eqref{eq:bec}) and six strain JVPs for the row term (Eq.~\eqref{eq:strain}), assembling $\Zpair$.}
\label{fig:method}
\end{figure*}

We present ERStream, a GPU execution engine for charge-aware equivariant MLIPs, applied to DPA4C-ER~\cite{dpa4clr2026} and new one- and two-layer MACE-ER models (MACE-ER-T1 and MACE-ER-T2), where ER denotes extended range and electrical response (Fig.~\ref{fig:method}).
ERStream treats the global solve as an electrostatic barrier, discarding forward neural activations beforehand and rematerializing them tile by tile afterward.
Only two scalars per atom cross this barrier: the raw learned charge entering the solve and its adjoint returning from it.
This partition confines neural memory to one active tile per GPU, makes the tile the unit of multi-GPU distribution, and returns local forces together with the pair-response tensor underlying the BECs for both strictly local and multi-hop models.
Both the trained parameters and the electrostatic formulation remain unchanged; in mixed precision, forces agree with a double-precision (FP64) evaluation of the same checkpoint to within $8.5\times10^{-5}$~eV\,\AA$^{-1}$ across the tested water and SrTiO$_3$ benchmarks.
We use compact architectures rather than large foundation potentials~\cite{macemp0, wood2025uma, dpa4} because distillation already produces compact short-range potentials that run more than tenfold faster~\cite{gardner_distillation_2025, amin_distilling_2026}. At the same time, our learned charges supply long-range electrostatics and electrical response.

Evaluating long-range forces and the pair-response tensor at every step, ERStream reaches about 83 million water atoms with MACE-ER-T2 and 300 million with MACE-ER-T1 on one NVIDIA B200 (180~GB).
On one consumer-grade NVIDIA GeForce RTX~5090 (32~GB), MACE-ER-T1 reaches 81 million water atoms, while DPA4C-ER reaches 40 million SrTiO$_3$ atoms and a bulk-water step time down to $9.49\times10^{-8}$~s/step/atom (at 5.2 million atoms).
Across eight NVIDIA B200 GPUs on one node, evaluating forces and the pair-response tensor achieves a 7.8-fold speed-up for a 2.1-million-atom periodic solvated protein.
ERStream also handles open boundary conditions for isolated systems, demonstrated on a 10-million-atom water droplet.
Trained on energies and forces alone, these models reproduce reference first-principles electrical observables, from the longitudinal--transverse optical (LO--TO) phonon splitting of SrTiO$_3$ to molecular infrared intensities.

\section{Results}

\subsection{ERStream execution engine}
\label{sec:engine}

For atomic positions $\mathbf r$ in a simulation cell $\mathbf h$, the local network $\phi$ predicts a site energy $E_i$ and a raw learned charge $\tq_i$ for each atom $i$. The site energies sum to the short-range energy $E^{\text{SR}}$, while the neutrality projection $\Pi_Q$ maps the raw charges $\qraw$ onto charges $\mathbf q$ with prescribed total charge $Q$. These projected charges determine the long-range electrostatic energy $E^{\text{LR}}$, giving the total energy $E=E^{\text{SR}}+E^{\text{LR}}$. The global solve evaluates $E^{\text{LR}}$ and its derivatives across all atoms using smooth particle-mesh Ewald (SPME) for periodic cells and a free-space solver for isolated systems. It also returns the charge adjoint $\mathbf w$ with components $w_i=\partial E^{\text{LR}}/\partial\tq_i$, which equals the electrostatic potential at atom $i$ minus the system-wide mean potential owing to the total-charge projection $\Pi_Q$. By the chain rule, the energy gradient splits into two terms~\cite{dpa4clr2026},
\begin{equation}
\begin{aligned}
 \frac{\partial E}{\partial\mathbf r}&=\left.\frac{\partial E^{\text{LR}}}{\partial\mathbf r}\right|_{\mathbf q}+\VJP_{\phi}\big[(1,\mathbf w)\big],\\
 \VJP_{\phi}\big[(1,\mathbf w)\big]&=\sum_{i}\left(\frac{\partial E_i}{\partial\mathbf r}+w_i\,\frac{\partial\tq_i}{\partial\mathbf r}\right),
\end{aligned}
 \label{eq:factor}
\end{equation}
with $\mathbf F=-\partial E/\partial\mathbf r$. The first term is the electrostatic gradient at fixed charges, $-\mathbf F^{\text{LR}}$. The second term is a vector--Jacobian product ($\VJP_{\phi}$), evaluated in one reverse-mode pass through the network with cotangent $(1,\mathbf w)$ applied to the site-energy and raw-charge outputs. Because $\mathbf w$ becomes available only after the global solve, this stage acts as an electrostatic barrier between the forward and reverse passes of the network. Conventional reverse-mode automatic differentiation retains the forward computational graph and all intermediate neural activations ($d\sim10^{4}$ scalar values per atom in our equivariant models) across this barrier to backpropagate gradients. For an $N$-atom system, this creates an $O(Nd)$ activation memory wall that worsens with message passing, because $T$ interaction layers with cutoff $r_c$ expand the receptive field of each learned charge to $T\,r_c$. Evaluating the electrical response additionally requires differentiating each learned charge with respect to all atomic positions within this $T\,r_c$ neighborhood.

Evaluated this way, the unoptimized MACE-ER baseline stores hundreds of kilobytes per atom and runs out of memory at 24,000 water atoms on one NVIDIA GeForce RTX~5090 (32~GB). To remove this activation memory wall, ERStream partitions each MD step at the electrostatic barrier. ERStream groups atoms into spatial tiles of $B$ atoms and evaluates the local network sequentially over all tiles (a sweep). Figure~\ref{fig:method}a illustrates the resulting three-stage step. First, the forward sweep computes and stores the site energies and raw charges for each tile and immediately discards its intermediate activations. Second, the global solve evaluates $E^{\text{LR}}$, the fixed-charge electrostatic forces, and the charge adjoint $\mathbf w$. Third, the reverse sweep rebuilds each tile from the atomic coordinates and evaluates Eq.~\eqref{eq:factor} together with the pair-response tensor $\Zpair$, which converts to the BECs via one global scaling factor (see Methods). Only two FP64 scalars per atom cross this electrostatic barrier, namely the raw charge entering the solve and its adjoint leaving it (8~bytes per atom each). As shown in Fig.~\ref{fig:method}b, all tiles reuse a single activation workspace of peak size $W(B)$, so only $O(N)$ per-atom arrays remain resident in GPU memory.

Figure~\ref{fig:method}c,d illustrates how a single MACE-ER-T2 tile is rebuilt and differentiated. Each tile carries a surrounding halo of thickness $T\,r_c$ containing all neighboring atoms that send messages to its owned atoms. Because neighboring tiles own halo atoms, the reverse sweep accumulates any force and pair-response contributions generated on halo atoms onto their owning tiles. On each rebuilt tile, one reverse-mode pass returns the local forces and stress (Fig.~\ref{fig:method}d). For MACE-ER-T2, explicitly storing the charge Jacobian $\partial\qraw/\partial\mathbf r$ would scale quadratically with the two-hop neighborhood size, so ERStream evaluates $\Zpair$ without materializing this Jacobian. Six forward-mode strain Jacobian--vector products (JVPs) through the rebuilt tile yield the row term of $\Zpair$, and one centered evaluation yields the column term (Fig.~\ref{fig:method}d; see Methods).

Algorithm~1 summarizes one complete ERStream step, pairing the shared forward sweep and global solve with backbone-specific derivative passes for MACE-ER-T2, MACE-ER-T1, and DPA4C-ER. Each derivative pass releases its computational graph before the next begins. All neural activations therefore remain confined to one tile, decoupling the system-wide $O(N)$ memory footprint from the feature dimension $d$ and interaction depth $T$.

\begin{algorithm*}[!t]
\footnotesize
\caption{\textbf{Algorithm 1. The ERStream step.} $O_t$ and $H_t$ are the owned atoms and the halo of tile $t$, $\mathbf a_i=\partial E_i/\partial\mathbf D_i$ and $\mathbf b_i=\partial\tq_i/\partial\mathbf D_i$ the cotangents with respect to the DPA4C descriptor $\mathbf D_i$ of atom $i$, $\mathbf e_q$ the unit cotangent on the charge output and $\epsilon_{\alpha\beta}$ the six strain components.}
\label{alg:step}
\SetKwInOut{Input}{Input}\SetKwInOut{Output}{Output}
\Input{positions $\mathbf r$, cell $\mathbf h$, total charge $Q$, tiles $t=1,\dots,n_B$}
\Output{$E$, $\mathbf F$, $\boldsymbol\sigma$, $\mathbf q$, $\Zpair$}
\ForEach{tile $t$}{
 $( E_i,\tq_i)_{i\in O_t}\leftarrow\phi(\mathbf r_{O_t\cup H_t})$\tcp*[r]{forward; activations discarded}
}
$\mathbf q\leftarrow\Pi_Q(\qraw)$;\quad $(E^{\text{LR}},\mathbf F^{\text{LR}},\boldsymbol\sigma^{\text{LR}},\partial E^{\text{LR}}/\partial\mathbf q)\leftarrow\mathrm{SPME}(\mathbf r,\mathbf h,\mathbf q)$;\quad $\mathbf w\leftarrow \mathbf P^{\top}\,\partial E^{\text{LR}}/\partial\mathbf q$\tcp*[r]{global solve, FP64}
\ForEach{tile $t$}{
 rebuild tile $t$ from $\mathbf r$\;
 \uIf{MACE-ER-T2}{
  $(\mathbf F,\boldsymbol\sigma)\mathrel{+}=\VJP_{\phi}[(1,\mathbf w)]$;\quad $\Zpair{}_{,i}\mathrel{+}=\mathrm{JVP}_{\phi}[\partial\tq_i/\partial\epsilon_{\alpha\beta}]$\tcp*[r]{column term and six strain JVPs (row term)}
 }
 \uElseIf{MACE-ER-T1}{
  $(\mathbf F,\boldsymbol\sigma)\mathrel{+}=\VJP_{\phi}[(1,\mathbf w)]$;\quad $\Zpair\mathrel{+}=\mathbf r\otimes\VJP_{\phi}[\mathbf e_q]$\tcp*[r]{shared tile graph}
 }
 \Else{
  $\mathbf u_i\leftarrow\mathbf a_i+w_i\mathbf b_i$,\ $\mathbf v_i\leftarrow\mathbf b_i$;\quad $(\mathbf F,\boldsymbol\sigma,\Zpair)\mathrel{+}=\textsc{EdgeKernel}(\mathbf u,\mathbf v,\partial\mathbf D/\partial\mathbf r)$\tcp*[r]{DPA4C-ER; one traversal}
 }
 release $t$\;
}
$E\leftarrow\sum_i E_i+E^{\text{LR}}$;\quad $\mathbf F\mathrel{+}=\mathbf F^{\text{LR}}$;\quad $\boldsymbol\sigma\mathrel{+}=\boldsymbol\sigma^{\text{LR}}$;\quad $\Zpair\mathrel{+}=\mathbf q\,\mathbf I$\tcp*[r]{assemble}
\end{algorithm*}

\subsection{Memory floor and multi-GPU scaling}
\label{sec:capacity}

After partitioning at the electrostatic barrier, the persistent GPU memory per atom approaches a floor (in bytes per atom),
\begin{equation}
 m_{\text{floor}}=m_{\text{graph}}+8\kappa+168+O(1/N),
 \label{eq:floor}
\end{equation}
where $m_{\text{graph}}$ is the persistent neighbor-structure storage in bytes per atom and $\kappa$ is the number of FP64 SPME mesh points per atom. The constant 168~bytes per atom stores the persistent FP64 per-atom arrays: positions and velocities (48~bytes), raw charges and adjoints (16~bytes), and projected charges, forces, and nine-component pair-response tensors (104~bytes). The available device memory $M_{\text{dev}}$ then bounds the reachable system size as
\begin{equation}
 N_{\max}\le\frac{M_{\text{dev}}-W(B)-M_{0}}{m_{\text{floor}}},
 \label{eq:nmax}
\end{equation}
where $W(B)$ is the single-tile activation workspace and $M_{0}$ is the fixed context and library allocation. By contrast, an unpartitioned whole-system evaluation adds an $O(d)$ activation term to the denominator.

\begin{figure*}[!t]
\centering
\includegraphics[width=\textwidth]{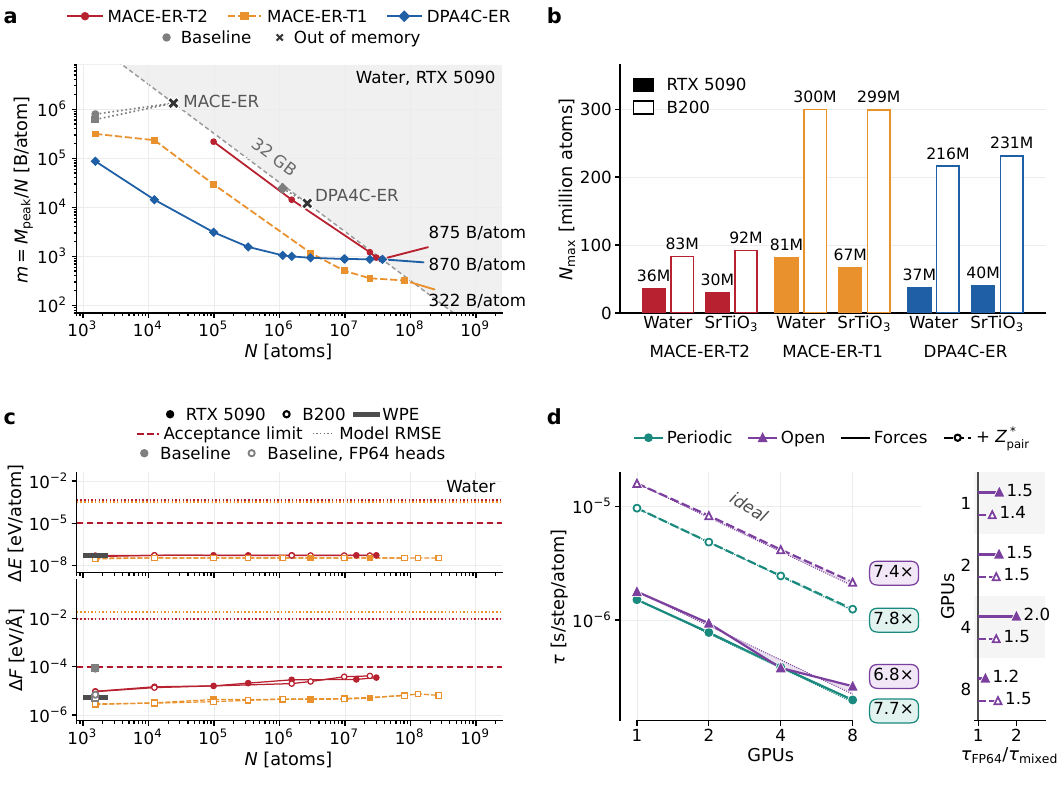}
\caption{\textbf{Memory, capacity, fidelity and multi-GPU scaling.} \textbf{a}, Peak memory per atom $m=M_{\text{peak}}/N$ for water on one NVIDIA GeForce RTX~5090 (32~GB dashed limit, exceeded region shaded). Model colors (MACE-ER-T2 red, MACE-ER-T1 orange, DPA4C-ER blue) apply across \textbf{a}--\textbf{c}. Gray markers show unoptimized baselines and crosses mark out-of-memory limits. \textbf{b}, Largest executed system $N_{\max}$ with $\Zpair$ at every step for water and SrTiO$_3$ on one RTX~5090 (filled) and one NVIDIA B200 (open). \textbf{c}, Maximum per-atom energy ($\Delta E$, top) and force ($\Delta F$, bottom) deviations of MACE-ER-T2 and MACE-ER-T1 in water relative to an FP64 evaluation of the same checkpoint at identical coordinates (GPU fill as in \textbf{b}). Dashed lines mark acceptance limits ($10^{-5}$~eV/atom, $10^{-4}$~eV\,\AA$^{-1}$) and dotted lines mark DFT RMSEs. Gray markers at $N=1{,}536$ show the FP32 baseline (filled) and FP32 with FP64 heads (open). 
The gray bar shows the deviation of qNEP evaluated with the Wisevolve Potential Engine (WPE)~\cite{wpe2026} from the same model evaluated with native GPUMD~\cite{gpumd2022}, both in FP32 and measured in this work. \textbf{d}, Mixed-precision step time per atom $\tau$ on one 8-GPU B200 node with distributed global solves (left) and droplet speed-up $\tau_{\mathrm{FP64}}/\tau_{\text{mixed}}$ (right). Periodic chignolin in water (teal circles, BioMol MACE-ER-T2, 2,148,768 atoms) and an open-boundary droplet (purple triangles, two-head MACE-ER-T2 water, 1,000,002 atoms) are shown without (solid, filled, Forces) and with (dashed, open, $+\Zpair$) the pair-response tensor. Dotted lines show ideal scaling and boxed numbers give 8-GPU speed-ups.}
\label{fig:scaling}
\label{fig:fidelity}
\end{figure*}

We verified this bound against the peak device memory for bulk water on one NVIDIA GeForce RTX~5090 (32~GB).
Figure~\ref{fig:scaling}a plots the peak memory per atom against system size. At their largest executed sizes, peak memory falls to 875~bytes per atom for MACE-ER-T2 (36 million atoms), 322~bytes per atom for MACE-ER-T1 (81 million atoms), and 870~bytes per atom for DPA4C-ER (37 million atoms). In contrast, the unoptimized MACE-ER baselines require hundreds of kilobytes per atom before running out of memory at 24,000 atoms, whereas the baseline DPA4C-ER evaluator~\cite{dpa4clr2026} requires 24.4~kB per atom (27.3~GB at 1.12 million atoms; SI Sec.~5). Across all three ERStream models, the measured per-atom memory converges toward the linear floor of Eq.~\eqref{eq:floor} as $N$ increases.

Figure~\ref{fig:scaling}b shows the largest executed system size $N_{\max}$ with $\Zpair$ evaluated at every step across models, GPUs, and materials (SI Sec.~5). In bulk water, MACE-ER-T2, MACE-ER-T1, and DPA4C-ER reach 36, 81, and 37 million atoms on one RTX~5090 (32~GB), and 83, 300, and 216 million atoms on one NVIDIA B200 (180~GB), respectively. Because the unoptimized MACE-ER baseline passes 12,288 water atoms on the RTX~5090 before failing at 24,000 atoms, ERStream increases single-GPU capacity by 1,500- to 2,900-fold for MACE-ER-T2 and 3,400- to 6,600-fold for MACE-ER-T1 (and 14- to 33-fold for DPA4C-ER; SI Sec.~5). Table~\ref{tab:capacity}a lists $N_{\max}$ and the per-atom step time $\tau$ for both water and SrTiO$_3$, where MACE-ER-T1 reaches around 300 million atoms on one B200.

To isolate numerical errors from tile partitioning and mixed-precision arithmetic, we compared the final configuration of each benchmark against an FP64 evaluation of the same checkpoint at identical coordinates. When executed entirely in FP64, tiled evaluation reproduces whole-system forces to within $6\times10^{-15}$~eV\,\AA$^{-1}$, matching floating-point rounding precision. Figure~\ref{fig:scaling}c shows the maximum per-atom energy ($\Delta E$) and force-component ($\Delta F$) deviations of mixed-precision ERStream relative to FP64 in water. Across 1,536 to 270 million water atoms, $\Delta F$ remains below $5.6\times10^{-5}$~eV\,\AA$^{-1}$, and across all tested water and SrTiO$_3$ systems it stays within $8.5\times10^{-5}$~eV\,\AA$^{-1}$ (SI Sec.~4). Every benchmark stays below our $10^{-4}$~eV\,\AA$^{-1}$ acceptance limit and remains 230- to 12,000-fold smaller than the corresponding model force RMSEs against DFT.

\begin{figure*}[!t]
\centering
\includegraphics[width=\textwidth]{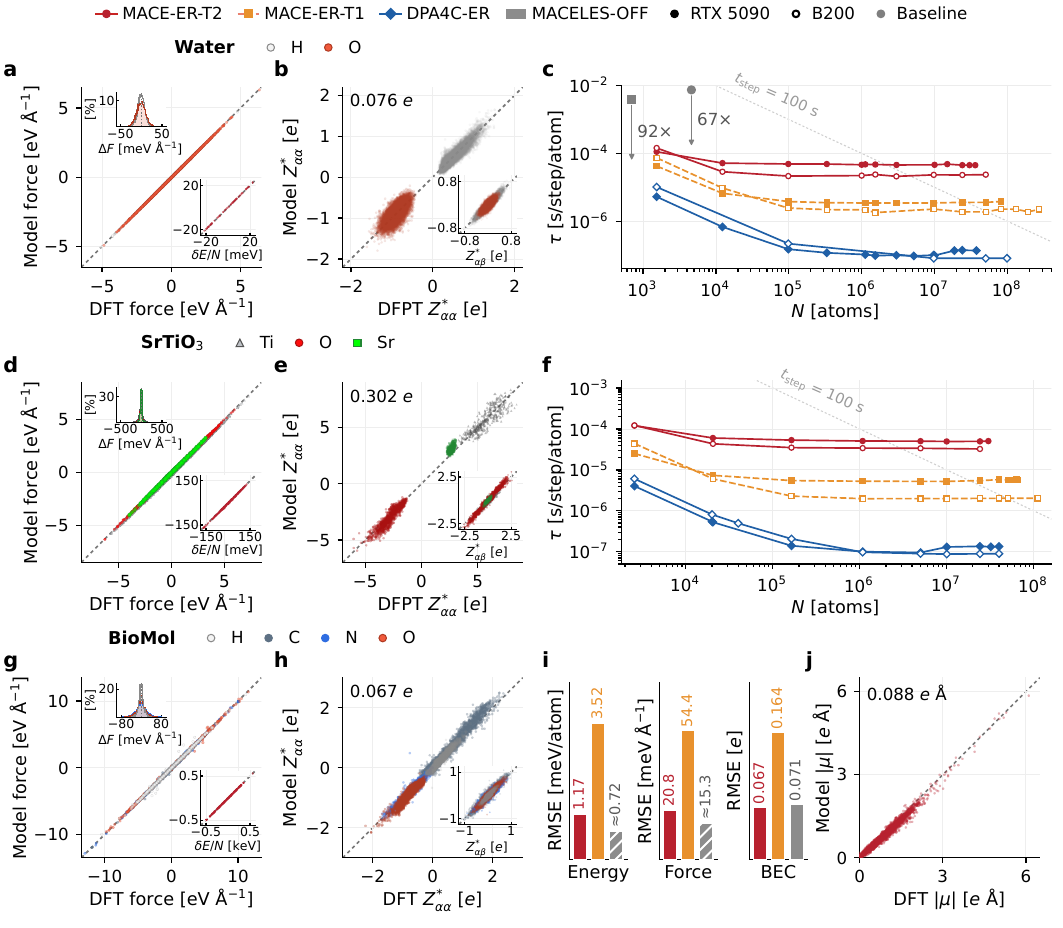}
\caption{\textbf{Model accuracy and performance.} \textbf{a}--\textbf{c}, Water. \textbf{d}--\textbf{f}, SrTiO$_3$. \textbf{a},\textbf{d}, MACE-ER-T2 forces against DFT. \textbf{b},\textbf{e}, MACE-ER-T2 diagonal Born effective charges $Z^{*}_{\alpha\alpha}$ against DFPT. \textbf{c},\textbf{f}, Step time per atom $\tau$ versus $N$ on one NVIDIA GeForce RTX~5090 (filled) and one NVIDIA B200 (open) for MACE-ER-T2 (red), MACE-ER-T1 (orange) and DPA4C-ER (blue), with dotted $t_{\text{step}}=100$~s lines. Gray markers and arrows in \textbf{c} show unoptimized baselines and speed-ups. \textbf{g}--\textbf{j}, BioMol. \textbf{g}, MACE-ER-T2 forces on the 11,037 test structures. \textbf{h}, $Z^{*}_{\alpha\alpha}$ on the 368-structure response pool. \textbf{i}, RMSE bars for MACE-ER-T2 (red), MACE-ER-T1 (orange) and MACELES-OFF (gray). The hatched MACELES-OFF energy and force bars combine the per-subset values of Ref.~\cite{les2025} with the subset composition of our test set. \textbf{j}, Dipole magnitudes $|\mu|$ of MACE-ER-T2 against DFT for 10,948 test structures with dipole labels. Force panels include species-resolved $\Delta F$ histograms (upper left) and $\delta E/N$ parity relative to the DFT test mean (lower right, meV in \textbf{a},\textbf{d}, keV in \textbf{g}). Born-charge panels include off-diagonal $Z^{*}_{\alpha\beta}$ insets and nine-component RMSEs (energy and force RMSEs in Table~\ref{tab:accuracy}). Models are trained on energies and forces alone, with one fitted scaling factor per periodic model and none for molecular models.}
\label{fig:accuracy}
\end{figure*}

\begin{table*}[!t]
\caption{\textbf{Capacity, speed and accuracy on one GPU.} \textbf{a}, $N_{\max}$ is the largest executed system and $\tau$ the time per atom-step of the largest benchmark on one NVIDIA GeForce RTX~5090 (32~GB) or one NVIDIA B200 (180~GB), 20 warm-up and 60 timed NVE steps with an FP64 end-point check. \textbf{b}, Test RMSEs against DFT and DFPT, shaded from green (low) to red (high). Water uses 50 test structures of 192 atoms and SrTiO$_3$ 1,445. Periodic Born charges carry one scaling factor per model, fitted on 65 water or eight SrTiO$_3$ structures and evaluated on 100 water or 22 held-out SrTiO$_3$ structures. BioMol Born charges are molecular and use no fitted scaling factor.}
\label{tab:capacity}
\label{tab:accuracy}
\begin{minipage}[t]{0.525\textwidth}\vspace{0pt}
\begin{ruledtabular}
\begin{tabular}{llcrcc}
\multicolumn{6}{l}{\textbf{a} Capacity and speed}\\
\colrule
 & & \multicolumn{2}{c}{$N_{\max}$} & \multicolumn{2}{c}{$\tau$ (s/step/atom)}\\
\cline{3-4}\cline{5-6}
System & Model & RTX~5090 & B200 & RTX~5090 & B200\\
\colrule
Water & MACE-ER-T2 & 36M & 83M & $4.52\times10^{-5}$ & $2.32\times10^{-5}$\\
 & MACE-ER-T1 & 81M & 300M & $3.82\times10^{-6}$ & $2.23\times10^{-6}$\\
 & DPA4C-ER & 37M & 216M & $1.37\times10^{-7}$ & $8.12\times10^{-8}$\\
 \colrule
SrTiO$_3$ & MACE-ER-T2 & 30M & 92M & $4.98\times10^{-5}$ & $3.25\times10^{-5}$\\
 & MACE-ER-T1 & 67M & 299M & $5.68\times10^{-6}$ & $2.02\times10^{-6}$\\
 & DPA4C-ER & 40M & 231M & $1.31\times10^{-7}$ & $8.94\times10^{-8}$\\
\end{tabular}
\end{ruledtabular}
\end{minipage}\hfill
\begin{minipage}[t]{0.45\textwidth}\vspace{0pt}
\begin{ruledtabular}
\begin{tabular}{llrrr}
\multicolumn{5}{l}{\textbf{b} Accuracy (RMSE)}\\
\colrule
 & & $E$ & $F$ & BEC\\
System & Model & (meV/atom) & (meV/\AA) & ($e$)\\
\colrule
Water & MACE-ER-T2 & \CE{0.440} & \CF{9.71} & \CZ{0.076}\\
 & MACE-ER-T1 & \CE{0.328} & \CF{18.80} & \CZ{0.056}\\
 & DPA4C-ER & \CE{0.293} & \CF{22.11} & \CZ{0.047}\\
\colrule
SrTiO$_3$ & MACE-ER-T2 & \CE{0.700} & \CF{27.99} & \CZ{0.302}\\
 & MACE-ER-T1 & \CE{4.01} & \CF{49.47} & \CZ{0.473}\\
 & DPA4C-ER & \CE{3.10} & \CF{95.02} & \CZ{0.351}\\
 \colrule
BioMol & MACE-ER-T2 & \CE{1.17} & \CF{20.78} & \CZ{0.067}\\
 & MACE-ER-T1 & \CE{3.52} & \CF{54.43} & \CZ{0.164}\\
\end{tabular}
\end{ruledtabular}
\end{minipage}
\end{table*}

Because each tile depends only on atomic coordinates within its local halo, ERStream distributes disjoint tile subsets across GPUs without exchanging intermediate neural activations, communicating only compact per-atom vectors and the electrostatic mesh at the global solve (see Methods). Figure~\ref{fig:scaling}d shows the mixed-precision step time per atom $\tau$ on one to eight NVIDIA B200 GPUs for a 2.1-million-atom periodic solvated chignolin box (BioMol MACE-ER-T2) and a 1.0-million-atom open-boundary water droplet (two-head MACE-ER-T2). For forces alone, scaling to eight GPUs accelerates the step by 7.7-fold for chignolin (96\% parallel efficiency) and 6.8-fold for the droplet (85\% efficiency). Evaluating $\Zpair$ at every step raises these eight-GPU speed-ups to 7.8-fold (98\% efficiency) and 7.4-fold (93\% efficiency), because the response passes add tile-local compute without increasing the serial time of the global solve. Distributing the global electrostatic solve across GPUs keeps the unparallelized serial time $t_{\mathrm G}$ small (see Methods, Eq.~\eqref{eq:strong}). The right panel of Fig.~\ref{fig:scaling}d compares mixed-precision and FP64 step times for the droplet across one to eight GPUs. Mixed precision accelerates execution by 1.2- to 2.0-fold for forces alone and by 1.4- to 1.5-fold when $\Zpair$ is evaluated.

\subsection{Model accuracy and performance}
\label{sec:MLIP-ER}

To evaluate accuracy and throughput, we benchmarked our models against DFT and density-functional perturbation theory (DFPT) across two periodic systems (bulk water and SrTiO$_3$) and the BioMol dataset of isolated molecules and clusters. 
We trained MACE-ER-T2 and MACE-ER-T1 for all three systems and DPA4C-ER for the two periodic systems, using only energies and forces.
Table~\ref{tab:accuracy}b lists the test-set root-mean-square errors (RMSEs) of each model against DFT energies and forces and DFPT (or finite-difference DFT) BECs. Across all three benchmarks, MACE-ER-T2 achieves the lowest force RMSEs, reducing errors by 43 to 62\% relative to MACE-ER-T1. Its second interaction layer expands the local receptive field from 6~\AA{} (one layer, $r_c=6$~\AA{}) to 8~\AA{} (two layers, $r_c=4$~\AA{}) and captures higher-order many-body correlations that a single layer misses.

Figure~\ref{fig:accuracy}a,b compares the MACE-ER-T2 forces and BECs against DFT and DFPT for bulk water. MACE-ER-T2 achieves a force RMSE of 9.7~meV\,\AA$^{-1}$ and, after calibrating one global scaling factor per periodic model, a BEC RMSE of 0.076~$e$ (compared with 0.056~$e$ for MACE-ER-T1 and 0.047~$e$ for DPA4C-ER). Figure~\ref{fig:accuracy}d,e shows the corresponding SrTiO$_3$ parity plots, where MACE-ER-T2 achieves RMSEs of 28.0~meV\,\AA$^{-1}$ in force and 0.302~$e$ in BEC. Compared with short-range counterparts sharing the same backbones and training data (SI Sec.~6), adding $E^{\text{LR}}$ (with staged fine-tuning for MACE-ER-T1) reduces the SrTiO$_3$ force RMSE by 21\% for both MACE-ER-T1 (63.0 to 49.5~meV\,\AA$^{-1}$) and DPA4C-ER (119.8 to 95.0~meV\,\AA$^{-1}$), and lowers the water force RMSE of MACE-ER-T1 by 15\% (22.2 to 18.8~meV\,\AA$^{-1}$).

The BioMol benchmark extends this evaluation to organic and biomolecular chemistry, spanning 198,696 training and 11,037 test C--H--O--N structures drawn from five subsets of the SPICE dataset~\cite{eastman2023spice} (DES370K monomers, DES370K dimers, dipeptides, solvated amino acids, and water clusters). Figure~\ref{fig:accuracy}g--j compares the forces, BECs, and dipole magnitudes of BioMol MACE-ER-T2 against DFT references and baseline models. On the 11,037 test structures, our compact 32-channel MACE-ER-T2 reaches an overall force RMSE of 20.8~meV\,\AA$^{-1}$ (18.6 to 29.1~meV\,\AA$^{-1}$ across subsets). For comparison, the 192-channel MACELES-OFF model achieves 11.7 to 25.3~meV\,\AA$^{-1}$ on the full SPICE subsets~\cite{les2025} (Fig.~\ref{fig:accuracy}i). On a 368-structure response pool spanning the same five subsets, MACE-ER-T2 predicts DFT BECs without any fitted scaling factor to an RMSE of 0.067~$e$, compared with 0.071~$e$ for MACELES-OFF and 0.164~$e$ for MACE-ER-T1. MACE-ER-T2 achieves lower BEC RMSEs than MACELES-OFF in four of the five subsets, all except dipeptides (SI Sec.~6). Across 10,948 test structures, predicted dipole magnitudes agree with DFT to an RMSE of 0.088~$e$\,\AA{} (Fig.~\ref{fig:accuracy}j). Thus, with one-sixth the channel width of MACELES-OFF, MACE-ER-T2 yields slightly higher force errors but comparable BEC accuracy.

Figure~\ref{fig:accuracy}c,f plots the step time per atom $\tau=t_{\text{step}}/N$ against system size $N$ on one RTX~5090 and one B200. Above $10^{5}$ atoms, $\tau$ plateaus for both MACE-ER models, because tile evaluation cost is independent of $N$ and the tile count grows linearly with $N$. In bulk water on one B200, $\tau$ plateaus at $2.3\times10^{-5}$~s/step/atom for MACE-ER-T2 and $2.2\times10^{-6}$~s/step/atom for MACE-ER-T1, with comparable plateaus in SrTiO$_3$ (Table~\ref{tab:capacity}a). Because the strictly local DPA4C backbone requires no message passing, DPA4C-ER reaches $8.1\times10^{-8}$~s/step/atom in water on the B200, 286-fold faster than MACE-ER-T2. Within each tile, MACE-ER runs cuEquivariance tensor-product kernels~\cite{cuequivariance} in 32-bit floating-point precision (FP32) with tabulated radial functions, whereas DPA4C-ER evaluates forces and pair-response contributions in a single fused edge kernel (SI Sec.~2). Relative to the unoptimized FP32 baselines on the same RTX~5090 with identical checkpoints (gray markers in Fig.~\ref{fig:accuracy}c), ERStream accelerates 1,536-atom water runs by 67-fold for MACE-ER-T2 and 92-fold for MACE-ER-T1.

\subsection{Electrical response}
\label{sec:response}

\begin{figure*}[!tp]
\centering
\includegraphics[width=\textwidth]{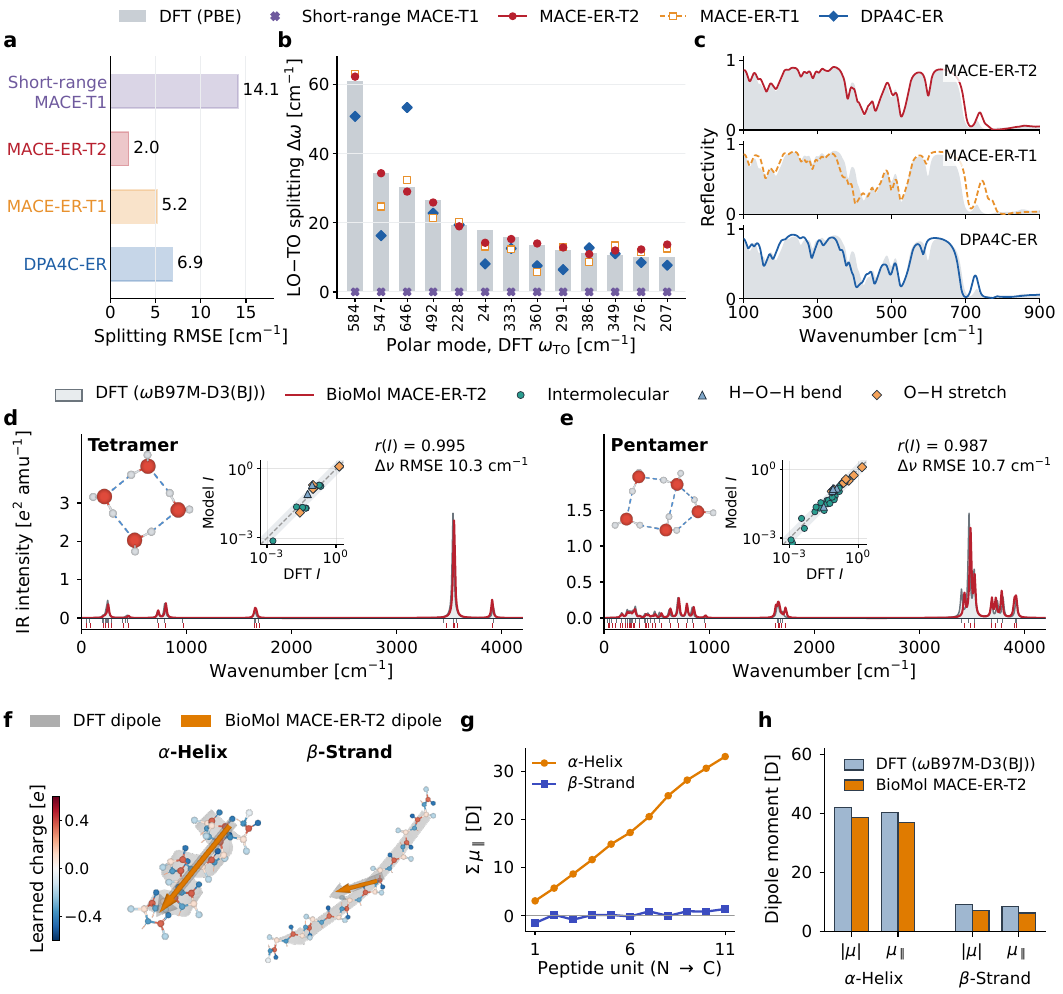}
\caption{\textbf{Electrical response from learned charges.} \textbf{a}--\textbf{c}, SrTiO$_3$ LO--TO splitting and infrared reflectivity of a held-out structure. \textbf{a}, Splitting RMSE across 57 $\Gamma$-point optical modes (short-range MACE-T1 omits electrostatics and yields zero splitting). \textbf{b}, Splitting $\Delta\omega=\omega_{\text{LO}}-\omega_{\text{TO}}$ of the 13 modes with PBE splitting $\ge 10$~cm$^{-1}$ (gray bars, PBE). \textbf{c}, Normal-incidence reflectivity of the three long-range models (curves) against PBE (gray fill). \textbf{d},\textbf{e}, Harmonic infrared spectra of the water tetramer (\textbf{d}) and pentamer (\textbf{e}) from BioMol MACE-ER-T2 (red lines) and $\omega$B97M-D3(BJ) DFT (gray fill) at identical DFT geometries, with bottom ticks marking mode wavenumbers. Log--log insets compare modes with non-zero DFT intensity $I$ (dashed parity, shaded two-fold band) for intermolecular ($<1{,}000$~cm$^{-1}$), H--O--H bend and O--H stretch modes, listing Pearson $r(I)$ and frequency RMSE ($\Delta\nu$). \textbf{f}, Out-of-distribution Ace--Ala$_{10}$--NMe $\alpha$-helix and $\beta$-strand colored by BioMol MACE-ER-T2 learned charges, with DFT (gray) and model (orange) dipole arrows. \textbf{g}, Cumulative axial dipole $\sum\mu_{\parallel}$ of BioMol MACE-ER-T2 from the N to the C terminus. \textbf{h}, Total $|\mu|$ and $\mu_{\parallel}$ against DFT. Panels \textbf{a}--\textbf{c} apply one calibrated scaling factor per periodic model, whereas \textbf{d}--\textbf{h} use unscaled molecular charges.}
\label{fig:response}
\end{figure*}

Beyond static BEC parity, we evaluated three physical observables governed by learned charges and their derivatives: the LO--TO phonon splitting and infrared reflectivity of SrTiO$_3$, harmonic infrared spectra of water clusters, and macrodipoles of an extended peptide. In the polar oxide SrTiO$_3$, the macroscopic electric field generated by polar lattice vibrations splits the longitudinal and transverse optical (LO and TO) phonons. Figure~\ref{fig:response}a--c compares the predicted LO--TO splitting and normal-incidence infrared reflectivity against PBE DFPT references. Each model combines its own force constants and calibrated BECs (one scaling factor per model) with the PBE DFPT electronic dielectric tensor of the same structure~\cite{kresse1996, kresse1999}. Across three held-out SrTiO$_3$ structures (SI Sec.~6), MACE-ER-T2 reproduces the LO--TO splitting with an RMSE of 1.7 to 2.9~cm$^{-1}$. For the structure in Fig.~\ref{fig:response}a,b, MACE-ER-T2 achieves a splitting RMSE of 2.0~cm$^{-1}$, compared with 5.2~cm$^{-1}$ for MACE-ER-T1, 6.9~cm$^{-1}$ for DPA4C-ER, and 14.1~cm$^{-1}$ for short-range MACE-T1 (which predicts zero splitting). Correspondingly, the normal-incidence reflectivity predicted by MACE-ER-T2 closely tracks the DFPT reference across the Reststrahlen bands (Fig.~\ref{fig:response}c).
The learned BECs therefore capture the coupling between polar phonons and the macroscopic electric field that governs the infrared optical response of SrTiO$_3$.

Figure~\ref{fig:response}d,e compares the harmonic infrared spectra of the water tetramer and pentamer computed with BioMol MACE-ER-T2 against $\omega$B97M-D3(BJ)/def2-TZVPPD DFT at identical DFT-optimized geometries~\cite{pyscf2020}. For the tetramer and pentamer, MACE-ER-T2 reproduces the harmonic frequencies to RMSEs of 10.3 and 10.7~cm$^{-1}$ and correlates with DFT mode intensities at Pearson $r=0.995$ and 0.987, respectively. 
Without any fitted scaling factor, the learned charge response thus reproduces the mode-resolved dipole derivatives that govern the vibrational infrared intensities of hydrogen-bonded water clusters.

As an out-of-distribution test beyond the dipeptide size limit of the BioMol training set, Fig.~\ref{fig:response}f--h compares the learned charges, cumulative axial dipoles, and total dipoles of the decapeptide Ace--Ala$_{10}$--NMe from BioMol MACE-ER-T2 against DFT.
In the $\alpha$-helix, peptide-unit dipoles align constructively along the chain axis (Fig.~\ref{fig:response}g) to produce a macrodipole of 38.5~D, compared with 42.0~D from DFT (Fig.~\ref{fig:response}h). In the $\beta$-strand, alternating peptide-unit dipoles largely cancel, yielding 7.0~D compared with 9.1~D from DFT. The learned charges thus distinguish the two secondary structures by their collective dipole.

\subsection{Protein dynamics in explicit water}
\label{sec:protein}

To test ERStream on a solvated biomolecule where charged side chains, salt bridges, and water polarization couple electrostatically, we simulated all-atom MD of the ten-residue $\beta$-hairpin chignolin (sequence GYDPETGTWG, PDB 1UAO)~\cite{honda2004} in explicit water. Each periodic box contains the peptide (net charge $-2$~$e$ at pH~5.5), two NH$_4^{+}$ counterions for charge neutrality, and 27,692 water molecules (83,224 atoms), matching the 277~K temperature and 2~mM peptide concentration of the nuclear magnetic resonance (NMR) measurements (BMRB 5694). Reference DFT scans of ammonium--carboxylate pairs confirm no spurious proton transfer by BioMol MACE-ER-T2, and all ions retain their charge states throughout production MD (SI Sec.~7). We propagated four 1~ns replicas from distinct NMR ensemble structures with BioMol MACE-ER-T2 (about 1~ns per day per replica on two B200 GPUs) and repeated the same four trajectories with Amber ff14SB~\cite{maier2015} and TIP3P water~\cite{jorgensen1983} as a classical baseline.

\begin{figure*}[!tp]
\centering
\includegraphics[width=\textwidth]{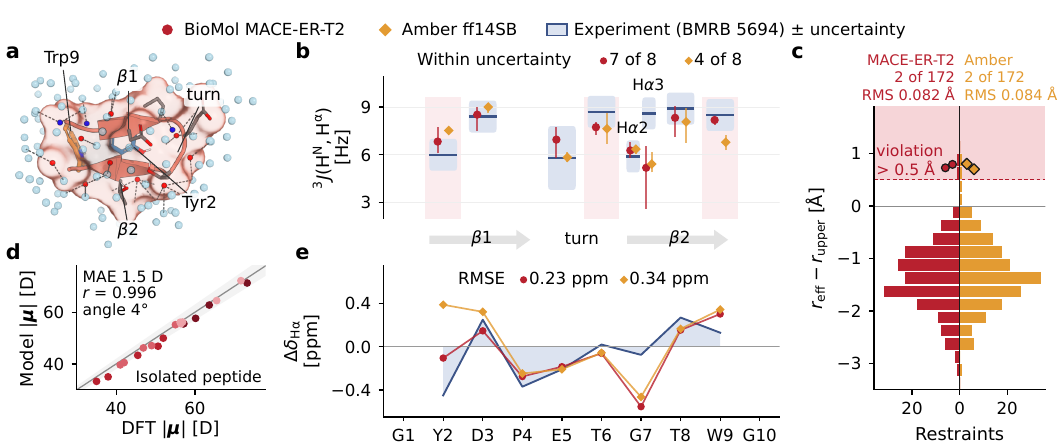}\\[1.0mm]
\includegraphics[width=\textwidth]{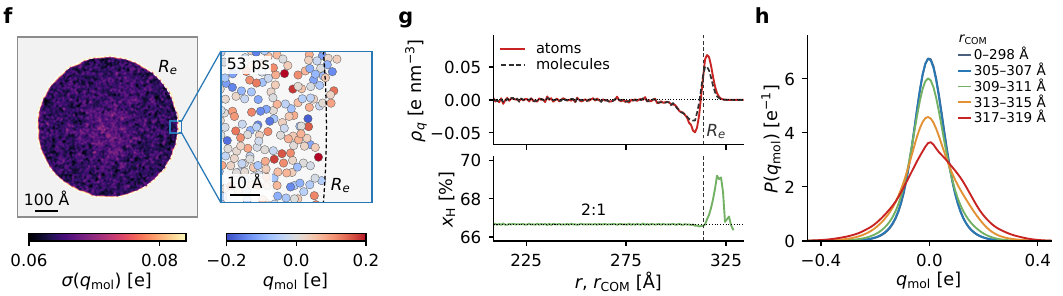}\\[2.2mm]
\includegraphics[width=\textwidth]{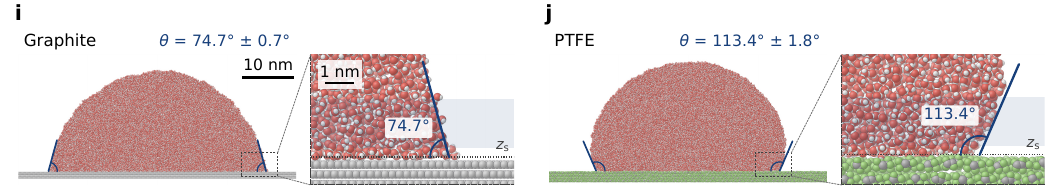}
\caption{\textbf{Protein dynamics and water at interfaces.} \textbf{a}--\textbf{e}, Chignolin in water at 277~K (four 1~ns replicas) comparing BioMol MACE-ER-T2 (red circles), Amber ff14SB (orange diamonds) and NMR experiment (BMRB 5694, navy). \textbf{a}, BioMol MACE-ER-T2 structure at 1~ns with nearest water molecules. \textbf{b}, Backbone couplings $^{3}J(\mathrm{H^N},\mathrm{H^\alpha})$ showing replica means (markers), ranges (bars), experiment $\pm$ uncertainty (navy lines, blue boxes) and couplings matched only by MACE-ER-T2 (pink columns). \textbf{c}, NOE distance excesses $r_{\text{eff}}-r_{\text{upper}}$ across 172 restraints (MACE-ER-T2 left, Amber right), with violations $>0.5$~\AA{} shaded and marked alongside RMS violations. \textbf{d}, Isolated-peptide dipole magnitudes $|\boldsymbol{\mu}|$ of 20 frames (DFT versus unscaled learned charges), shaded by replica, with a $\pm 5\%$ band. \textbf{e}, Secondary chemical shifts $\Delta\delta_{\mathrm{H}\alpha}$. \textbf{f}--\textbf{h}, Isolated 10-million-atom water droplet at 300~K (two-head MACE-ER-T2 bulk-water head, scaled charges, 1--53~ps average). \textbf{f}, RMS molecular charge $\sigma(q_{\text{mol}})$ in equatorial slab $|z|<10$~\AA{} (left) and surface snapshot at 53~ps colored by $q_{\text{mol}}$ (right). \textbf{g}, Radial charge density $\rho_q$ of atoms versus $r$ (solid) and molecules versus $r_{\text{COM}}$ (dashed, top), and hydrogen percentage $x_{\mathrm H}$ (bottom, dotted 2:1 line). Dashed lines in \textbf{f} and \textbf{g} mark the equimolar radius $R_e$. \textbf{h}, Distributions $P(q_{\text{mol}})$ across $r_{\text{COM}}$ shells. \textbf{i},\textbf{j}, Water nanodroplets on graphite (\textbf{i}) and PTFE (\textbf{j}) at 300~K with Interface MACE-ER, after 200~ps. Left, a 1.2~nm slice through the droplet axis, cut at $45^{\circ}$ to the chains for PTFE. Right, the magnified right contact line, showing the contact plane $z_{\mathrm s}$ (dotted), the fit band (shaded) and the contact angle $\theta$ (blue lines, mean $\pm$ s.d. over 190--200~ps).}
\label{fig:chignolin}
\label{fig:droplet}
\end{figure*}

Figure~\ref{fig:chignolin}a shows a MACE-ER-T2 replica after 1~ns. The $\beta$-hairpin remains folded in all four MACE-ER-T2 replicas, with a mean backbone root-mean-square deviation (RMSD) of 0.89~\AA{} from the first NMR structure. 
Figure~\ref{fig:chignolin}b compares the backbone $^{3}J(\mathrm{H^N},\mathrm{H^\alpha})$ couplings back-calculated via the Karplus relation~\cite{vogeli2007} against the NMR measurements. MACE-ER-T2 matches seven of the eight couplings within experimental uncertainty (RMSE 1.38~Hz), compared with four of eight for Amber ff14SB (RMSE 1.49~Hz). Hierarchical bootstrap resampling over replicas and 25~ps blocks yields overlapping 95\% confidence intervals for the coupling RMSEs (1.01--1.99~Hz for MACE-ER-T2 versus 1.27--1.74~Hz for Amber), showing that the two ensembles are statistically indistinguishable on this metric. 
The two ensembles differ most at Y2, T6, and W9, and both underestimate the G7 H$^{\alpha}$3 coupling of the turn.

Figure~\ref{fig:chignolin}c compares the effective interproton distances from both ensembles against the 172 deposited nuclear Overhauser effect (NOE) upper distance bounds. Both ensembles show two violations exceeding 0.5~\AA{} (both involving the W9 indole proton H$^{\varepsilon 3}$), with RMS violations of 0.082~\AA{} for MACE-ER-T2 and 0.084~\AA{} for Amber. Figure~\ref{fig:chignolin}e presents the H$^{\alpha}$ secondary chemical shifts predicted with SPARTA+~\cite{shen2010}. MACE-ER-T2 yields an H$^{\alpha}$ secondary-shift RMSE of 0.23~ppm (95\% bootstrap interval 0.17--0.29~ppm), matching the 0.25~ppm validation error of SPARTA+, whereas Amber ff14SB gives 0.34~ppm. 
Across these back-calculated NMR observables, the MACE-ER-T2 ensemble is consistent with the solution structure of chignolin and matches or improves upon the Amber ff14SB/TIP3P baseline. ERStream thus scales a quantum-chemistry-trained long-range potential to explicit-solvent protein trajectories at the experimental peptide concentration.

To validate the learned charges directly on trajectory-sampled peptide conformations, we extracted 20 chignolin frames without solvent and compared their isolated-peptide dipoles from BioMol MACE-ER-T2 against $\omega$B97M-D3(BJ)/def2-TZVPPD DFT (Fig.~\ref{fig:chignolin}d). Without any fitted scaling factor, the unscaled learned-charge dipoles agree with DFT to a mean absolute error of 1.5~D in magnitude ($r=0.996$) and a mean angular deviation of $4^{\circ}$.

\subsection{Water droplets and surface wetting}
\label{sec:droplet}

To simulate isolated systems without periodic images (e.g., droplets and nanoparticles), ERStream retains the same step partition and replaces only the periodic SPME stage with an open-boundary electrostatic solver. Instead of evaluating the erf-screened Coulomb sum of Eq.~\eqref{eq:freespace} directly at $O(N^{2})$ cost, ERStream splits the kernel into a short-ranged real-space pair sum and a smooth long-range term evaluated by zero-padded fast Fourier transform (FFT) convolution~\cite{hockney1988}. This split solver scales as $O(N\log N)$, reproduces the direct sum to within $5.3\times10^{-5}$~eV\,\AA$^{-1}$ in force at one million atoms (see SI Sec.~5), and distributes across GPUs alongside the spatial tiles.

We evaluated this open-boundary solver at scale by simulating an isolated 10-million-atom water droplet at 300~K using the bulk-water head of the two-head MACE-ER-T2 water model on eight B200 GPUs.
Following 63~ps of equilibration, we ran 60~ps of Nos\'e--Hoover production dynamics (removing evaporated molecules beyond 600~\AA{} once at 38~ps) and averaged radial charge profiles over 1 to 53~ps (SI Sec.~3). Below, we report projected learned charges scaled by the calibrated factor $|s|=5.5398$ and give radii relative to the equimolar radius $R_e=313.8$~\AA{}.

Figure~\ref{fig:droplet}f--h shows the equatorial RMS molecular charge and surface snapshot (\textbf{f}), radial atomic and molecular charge-density profiles in 1~\AA{} shells (\textbf{g}), and shell-resolved molecular charge distributions (\textbf{h}). The droplet develops a negative subsurface atomic charge-density layer reaching $-0.049$~$e$\,nm$^{-3}$ roughly 4~\AA{} inside $R_e$, beneath a hydrogen-enriched positive outer layer peaking at $+0.068$~$e$\,nm$^{-3}$. The molecular charge density exhibits the same double layer ($-0.031$~$e$\,nm$^{-3}$ subsurface minimum and $+0.050$~$e$\,nm$^{-3}$ surface peak; Fig.~\ref{fig:droplet}g), showing that whole surface molecules carry net positive charge above a negatively charged subsurface layer while the droplet interior remains neutral. This layer sequence agrees qualitatively with \citet{liangyang2025}, who simulated a 10,327-molecule droplet using a polarizable model trained explicitly on MP2 charges. ERStream thus resolves interfacial charge layering in isolated systems at the 10-million-atom scale without charge-label training.

For multi-element solid--liquid interfaces, we deployed a separately trained long-range model (Interface MACE-ER) without modifying the engine.
Interface MACE-ER is a one-layer, 64-channel model with code-specific readout heads trained on RPBE-D3(BJ) energies and forces of water, ice, solid surfaces, and interfaces~\cite{hammer1999rpbe, grimme2011d3bj} (see Methods and SI Sec.~1). We placed 558,234-molecule water nanodroplets on graphite (3.2 million atoms) and polytetrafluoroethylene (PTFE, 3.4 million atoms) and propagated each for 200~ps at 300~K on eight B200 GPUs, switching from periodic to slab electrostatics after 50~ps (SI Sec.~3). Figure~\ref{fig:droplet}i,j shows the final droplet profiles and contact lines at 200~ps. Averaged over 190 to 200~ps, the contact angles are $74.7^{\circ}\pm0.7^{\circ}$ on graphite and $113.4^{\circ}\pm1.8^{\circ}$ on PTFE. 
ERStream thus extends learned-charge MD to multi-element solid--liquid interfaces at the multi-million-atom scale.

\section{Discussion}

In conventional evaluations of charge-aware MLIPs, retaining the forward neural computation graph of every atom across the global electrostatic solve creates an $O(Nd)$ activation memory wall.
ERStream removes this bottleneck by discarding tile activations after the forward charge prediction and passing only one raw charge $\tq_i$ and one electrostatic adjoint $w_i$ per atom across the global electrostatic solve to evaluate local forces and the pair-response tensor $\Zpair$.
This forward--solve--reverse step decomposition confines the neural workspace to a single tile of $B$ atoms, so system-wide GPU memory scales with the persistent per-atom state $m_{\text{floor}}$ (Eq.~\eqref{eq:floor}) rather than with the feature dimension $d$ or interaction depth $T$.
Single-GPU capacity is therefore decoupled from the computational complexity of the equivariant backbone.
Across DPA4C-ER, MACE-ER-T1 and MACE-ER-T2, the step time varies by a factor of 286 on one NVIDIA B200. Yet, all three models accommodate 83 to 300 million water atoms with full pair-response evaluation at every step (Table~\ref{tab:capacity}).

Evaluating long-range electrostatics and BECs at hundred-million-atom scales brings collective electrical phenomena within reach of MLIP-based molecular dynamics.
In ferroelectric superlattices, polar vortices and skyrmions span tens of nanometers~\cite{yadav2016, das2019}, while at electrochemical interfaces, electric double layers couple electrode charge to the bulk electrolyte over comparable screening lengths~\cite{wang2025edl}.
Because the BEC tensors determine the forces induced by an applied electric field~\cite{stocco2025field}, simulations at this scale could follow polarization switching, ion migration and electrothermal heat transport under bias in nanoscale devices~\cite{aryana2022}.
In aqueous solution, large periodic boxes allow biomolecules to be simulated at experimental peptide concentrations and temperatures, as shown by our chignolin trajectories at 2~mM and 277~K, which are consistent with measured NMR observables (Fig.~\ref{fig:chignolin}).
With potentials trained for extreme pressures and temperatures, charge-aware simulations at this scale could also model superionic water in ice-giant interiors~\cite{cheng2021superionic} and predict the infrared spectra of astrophysical ices~\cite{trumbo2023europa} and planetary water plumes~\cite{villanueva2023enceladus} probed by space-based infrared spectroscopy.

Rematerializing tile activations across the global electrostatic solve evaluates the exact computational graph of the trained checkpoint without modifying model parameters or truncating long-range interactions.
In double precision (FP64), tiled execution reproduces whole-system force evaluations to within $6\times10^{-15}$~eV\,\AA$^{-1}$, matching floating-point rounding precision.
In mixed precision, where the equivariant backbone runs in FP32 while readouts, charge projection and the electrostatic solve remain in FP64, forces deviate by at most $8.5\times10^{-5}$~eV\,\AA$^{-1}$ from the FP64 reference across all water and SrTiO$_3$ benchmarks.
This numerical deviation is 230 to 12,000 times smaller than the corresponding force RMSEs of the models against DFT.

With memory bounded near the per-atom floor, practical simulations are governed by the trade-off between model accuracy and step time.
The nearest charge-aware engine at comparable system sizes is the Wisevolve Potential Engine (WPE) running qNEP \cite{wpe2026}, which evaluates forces and BECs from analytical derivatives of a compact local descriptor.
In our benchmarks on one RTX~5090, WPE executes qNEP with BECs for up to 8.9 million water atoms at $1.6\times10^{-8}$~s/step/atom, running six to nine times faster than DPA4C-ER, whereas our equivariant model designs prioritize higher force and response accuracy~\cite{dpa4clr2026}.
Trained on the same SCAN water dataset as qNEP, MACE-ER-T2 achieves a force RMSE of 17.8~meV\,\AA$^{-1}$, compared with 70 to 76~meV\,\AA$^{-1}$ for qNEP (SI Sec.~5).
The two engines thus target different regimes of the speed--accuracy frontier, while ERStream accommodates four to nine times more water atoms (36 to 81 million) on the same 32~GB GPU.

Within ERStream, the time per MD step similarly sets the accessible simulation timescale for each architecture.
The strictly local DPA4C-ER is the fastest of our three models at $8.1\times10^{-8}$~s/step/atom on one NVIDIA B200, delivering about 0.5~ns per day for one million atoms at a 0.5~fs time step.
At 216 million water atoms, this throughput drops to about 2.5~ps per day, meaning that hundred-million-atom runs currently serve as capacity benchmarks or short relaxations rather than long production trajectories.
At the high-accuracy end, MACE-ER-T2 retains full O(3) equivariant tensor products over a two-hop neighborhood and yields the lowest force RMSEs across all three benchmarks (Table~\ref{tab:accuracy}b), but runs 286 times slower than DPA4C-ER.
Most of this cost arises from evaluating the two-hop pair-response tensor $\Zpair$ via sequential forward-mode strain tangents, which lengthens each MACE-ER-T2 step six- to ninefold compared with evaluating energies and forces alone (Fig.~\ref{fig:scaling}d).
A microsecond trajectory of a million-atom solvated protein, the timescale on which fast-folding proteins fold~\cite{lindorff2011}, remains beyond single-node throughput.

Streaming spatial tiles with activation rematerialization across the global electrostatic solve requires two structural properties of the interatomic potential.
First, the site energy $E_i$ and raw learned charge $\tq_i$ of each atom must depend only on neighbors within a finite receptive field $T\,r_c$.
Second, the local neural network and the global electrostatic solver must couple exclusively through low-dimensional per-atom scalars, here $\tq_i$ in the forward pass and the electrostatic adjoint $w_i$ in the reverse pass.
Self-consistent charge-equilibration potentials such as the fourth-generation high-dimensional neural network potential~\cite{ko2021} and global-attention architectures~\cite{kreiman2025, frank2026efa, qu_recipe_2026} violate these locality conditions by coupling all atoms through a system-wide charge-equilibration matrix or all-to-all attention layers.
By contrast, we expect any long-range MLIP that pairs a finite-range local backbone with a global electrostatic stage coupled through a few scalars or vectors per atom to remain streamable at arbitrary feature dimension $d$ and finite interaction depth $T$.

For distributed execution across GPUs, ERStream adopts a different communication architecture from message-passing engines such as Tessera and DistMLIP~\cite{tessera2026, dispmlip}, which distribute deep short-range graph networks by exchanging intermediate node features across domain boundaries at every interaction layer.
To reach a 23~\AA{} receptive field without an explicit electrostatic term, Tessera stacks four message-passing layers and transfers 14~kB of hidden features per boundary atom at each layer.
Because ERStream delegates long-range interactions to the explicit electrostatic solve over learned charges, its equivariant backbones require only one or two interaction layers with a local reach of 6 to 8~\AA{} (SI Sec.~1).
Each GPU can therefore evaluate the halos of its tiles directly from atomic coordinates rather than exchanging intermediate layer features, reducing inter-GPU communication on a node to 72~bytes per atom per step outside the global solve (Eq.~\eqref{eq:exchange}, or 144~bytes with $\Zpair$). 
Extending ERStream across multiple nodes would replace the single-node position broadcast with spatial domain decomposition that exchanges only coordinates, forces and pair-response contributions within the 6 to 8~\AA{} boundary halo between neighboring nodes.
The global SPME or free-space electrostatic solve would then run on a distributed three-dimensional FFT mesh, while the local--global interface across the solve remains a single scalar charge $\tq_i$ and adjoint $w_i$ per atom.

Finally, we envision that replacing the O(3) tensor products inside each tile with faster equivariant operations that keep a finite receptive field and a scalar coupling to the solve, such as the SO(2) convolutions of eSEN and DPA4~\cite{esen2025, dpa4} accelerated by low-precision matrix-multiplication emulation~\cite{lu2026so2}, would directly reduce the tile evaluation time.
In parallel, multiple-time-step integration could evaluate the slower multi-hop forces and pair-response tensor at a larger outer time step~\cite{tuckerman1992}.

In summary, ERStream removes the activation memory wall of charge-aware MLIPs by decoupling tiled neural evaluation from the global electrostatic solve through activation rematerialization, without altering the trained potential.
This execution schedule evaluates long-range forces and the pair-response tensor at every step within a tile-bounded memory footprint on a single GPU and scales across multiple GPUs on a node.
With memory bounded near the per-atom floor, the wall-clock time per MD step now determines the accessible simulation timescales.
Combining this streaming execution architecture with faster equivariant kernels and algorithm--hardware co-design will further extend the timescale reach of device-scale charge-aware atomistic simulations.

\section*{Methods}

\method{Electrostatic barrier partition}
The total energy of a long-range MLIP with learned charges is the sum of short-range and long-range terms, $E=E^{\text{SR}}+E^{\text{LR}}$. The local network $\phi$ maps the atomic numbers $Z_j$ and positions $\mathbf r_j$ within the dependency neighborhood $\mathcal N_i$ of each atom $i$ to the site energy and the raw learned charge~\cite{dpa4clr2026},
\begin{equation}
 E^{\text{SR}}=\sum_i E_i,\qquad \tq_i=\phi_q(\{Z_j,\mathbf r_j\}_{j\in\mathcal N_i}),
 \label{eq:heads}
\end{equation}
where the dependency reach is $T\,r_c$ for MACE-ER with $T$ interaction layers and the descriptor cutoff $r_c$ for DPA4C-ER~\cite{les2025, dpa4clr2026}. The charges are learned from reference energies and forces without charge labels, as in Ref.~\cite{les2024}.
For a system of $N$ atoms, the raw charges are neutralized to the prescribed total charge $Q$ by the projection
\begin{equation}
 q_i=\Pi_Q(\tq)_i=\tq_i+\frac{Q-\sum_\ell\tq_\ell}{N},
 \label{eq:projection}
\end{equation}
whose Jacobian is the symmetric projector $\mathbf P=\mathbf I_N-\mathbf 1\mathbf 1^{\top}/N$, with $\mathbf I_N$ the $N\times N$ identity matrix and $\mathbf 1$ the $N$-vector of ones. Under periodic boundary conditions with cell matrix $\mathbf h$, the projected charges $\mathbf q$ interact through a Gaussian-screened Coulomb kernel evaluated by smooth particle-mesh Ewald (SPME)~\cite{pme1995} with splitting parameter $\alpha$, yielding the long-range energy $E^{\text{LR}}(\mathbf r,\mathbf h,\mathbf q)$.
The total force on atom $j$ comprises three terms~\cite{dpa4clr2026},
\begin{equation}
\begin{split}
 \mathbf F_j&=-\frac{\partial E^{\text{SR}}}{\partial\mathbf r_j}
 -\left.\frac{\partial E^{\text{LR}}}{\partial\mathbf r_j}\right|_{\mathbf q}
 -\sum_i w_i\,\frac{\partial\tq_i}{\partial\mathbf r_j}\\
 w_i&\equiv\frac{\partial E^{\text{LR}}}{\partial\tq_i}
\end{split}
 \label{eq:force}
\end{equation}
the short-range force, the electrostatic force at fixed charges and the local charge-response force. The charge adjoint is $\mathbf w=\mathbf P^{\top}\,\partial E^{\text{LR}}/\partial\mathbf q$ and incorporates the derivative of the neutrality projection in Eq.~\eqref{eq:projection}.
For periodic models, we evaluate the polarization and Born effective charges (BECs) using the rescaled charges $\tilde q_i=s\,q_i$, where $s$ is a signed calibration factor. The BEC of atom $i$ decomposes into a direct charge contribution and a charge-response contribution,
\begin{equation}
\begin{split}
 \mathbf Z^{*}_i&=\tilde q_i\,\mathbf I+\frac{s}{2}\sum_{j\neq i}\left(\mathbf r_{ij}\otimes\frac{\partial\tq_i}{\partial\mathbf r_{ij}}-\mathbf r_{ji}\otimes\frac{\partial\tq_j}{\partial\mathbf r_{ji}}\right)\\
 &=s\,\Zpair{}_{,i},
\end{split}
\label{eq:bec}
\end{equation}
where $\otimes$ denotes the outer product. The global charge projection cancels from the pairwise response term, which therefore acts on the raw charges. 
The tensor $\Zpair$ is the unscaled pair-response tensor that the engine returns at every step, with the projected charge $q_i$ in its direct term. Its pairwise term sums to zero over all atoms. $\Zpair$ therefore obeys the acoustic sum rule $\sum_i\Zpair{}_{,i}=Q\,\mathbf I$, which the FP64 evaluations of the test structures satisfy to $8\times10^{-14}$~$e$ for water and $6\times10^{-15}$~$e$ for SrTiO$_3$.
Following Ref.~\cite{dpa4clr2026}, we fit $s$ for each periodic model as the least-squares slope of the DFPT BECs against the pair-response tensor on a calibration set and keep it fixed for evaluation. We report its square, $\varepsilon_{\text{fit}}=s^{2}$, in SI Table~S18.

Equations~\eqref{eq:heads}--\eqref{eq:bec} define the three stages of one MD step (Fig.~\ref{fig:method}a and Algorithm~1),
\begin{equation}
\begin{aligned}
 (\mathbf r,\mathbf h)\ &\xrightarrow{\ \phi\ }\ (E^{\text{SR}},\qraw),\\
 (\mathbf r,\mathbf h,\qraw)\ &\xrightarrow{\ \Pi_Q,\ \mathrm{SPME},\ \mathbf P^{\top}\ }\ (E^{\text{LR}},\mathbf q,\mathbf F^{\text{LR}},\boldsymbol\sigma^{\text{LR}},\mathbf w),\\
 (\mathbf r,\mathbf h,\mathbf q,\mathbf w)\ &\xrightarrow{\ \VJP_{\phi},\ \partial\qraw/\partial\mathbf r\ }\ (\mathbf F,\boldsymbol\sigma,\Zpair),
\end{aligned}
\label{eq:chain}
\end{equation}
where $\mathbf F^{\text{LR}}$ and $\boldsymbol\sigma^{\text{LR}}$ are the force and stress at fixed charges. ERStream executes the first and third stages sequentially over spatial tiles, and the second stage remains the only system-wide computation. Across the partition between the local and global stages, ERStream passes only the raw learned charges $\qraw$ and the scalar adjoints $\mathbf w$, FP64 scalars of 8~bytes per atom each. Atomic geometry, neighbor structures, the FP64 reciprocal-space mesh and the output arrays remain system-sized, whereas every neural activation is confined to a single tile.

\method{Rematerialized tile streaming}
The partition at the electrostatic barrier confines every neural activation to one tile. Tiles are contiguous ranges of $B$ atoms in a fixed atom order. MACE-ER-T2 sorts atoms by the Morton key of their cell-list bin, MACE-ER-T1 uses the input configuration order, and DPA4C-ER schedules atoms in cell order. A tile owns one such range and alone writes the site energies, learned charges, and pair-response tensors of its atoms. Its halo is the complete dependency neighborhood of the owned atoms to depth $T\,r_c$, with periodic images entering through lattice translations of the edge vectors (Fig.~\ref{fig:method}c). Interaction layer $\ell$ of $T$ updates only the nodes within $(T-\ell)\,r_c$ of an owned atom.

For MACE-ER-T2, 
a tile holds its owned atoms and a halo of two shells: an inner shell of atoms within $r_c$ of an owned atom and an outer shell of atoms within $r_c$ of the inner shell (Fig.~\ref{fig:method}c). The first interaction layer reads both shells and updates the inner shell and the owned atoms. The second reads the inner shell and updates the owned atoms only. When the pair-response tensor is evaluated in a periodic cell, periodic images enter the tile as distinct nodes at $\mathbf r+\mathbf n\mathbf h$, with $\mathbf n$ an integer lattice vector. The neural layers run in FP32. Geometry, output heads, charges and the accumulation of forces and pair-response tensors use FP64, as does the electrostatic solve, which evaluates SPME with cuFFT. Geometry, neighbor data, charges, adjoints and outputs stay resident as system-sized buffers. The SPME mesh is a single FP64 array that holds the real-space grid and, after an in-place transform, its half-spectrum. 
The forward sweep evaluates the local network tile by tile, writes the short-range energies and raw learned charges of the owned atoms, and immediately discards all intermediate activations.
After the global solve, the reverse sweep rebuilds each tile's activations from the retained coordinates and neighbor structure, evaluates the derivatives in Eqs.~\eqref{eq:force} and \eqref{eq:bec}, and accumulates them into system-sized output arrays in FP64 before releasing the tile.
This schedule applies activation rematerialization~\cite{chen2016, jain2020} across the global electrostatic barrier (Fig.~\ref{fig:method}b).
The neural working set scales with $B$ and is independent of $N$, and the tiled sweeps return the same outputs as a single evaluation of the whole system, up to floating-point rounding and accumulation order.

For a compact tile whose owned atoms fill a cube of edge $L_B=(B/\rho)^{1/3}$ at number density $\rho$, the halo is the shell of thickness $T\,r_c$ around it and adds the halo factor
\begin{equation}
 h(B)=\Big(1+\frac{2\,T\,r_c}{L_B}\Big)^{3}-1,\qquad L_B=\Big(\frac{B}{\rho}\Big)^{1/3},
 \label{eq:halo}
\end{equation}
to the evaluated atoms. Tiles of other shapes have a larger surface and a larger halo. The forces cost $2\,[1+h(B)]$ forward evaluations per owned atom and step, one half of which is the cost of rematerialization, and the response passes add to this. The tile workspace is $W(B)=O\!\big(B\,[1+h(B)]\,d\big)$.
Larger tiles reduce both the halo factor $h(B)$ and the kernel-launch count while increasing the workspace $W(B)$. The tile size is chosen per device and model from this trade-off. A planner selects the number of tiles from the structure and the available device memory, and the user can also set it explicitly. Adopted values and ablations are given in SI Sec.~5 and 8.

\method{Local Jacobian contractions}
The reverse sweep contracts the local Jacobian $\partial\qraw/\partial\mathbf r$ of each rebuilt tile twice. The $(1,\mathbf w)$-weighted contraction gives the forces and virial in Eq.~\eqref{eq:force} in one reverse-mode pass for any interaction depth. The unweighted contraction gives the charge-response contribution in Eq.~\eqref{eq:bec}, and its evaluation depends on the dependency reach of the network (Algorithm, lines 7--13, and Fig.~\ref{fig:method}d).

For MACE-ER-T2, each learned charge depends on atoms two hops away. Its Jacobian couples all atoms of the dependency neighborhood, and storing it explicitly would scale quadratically with the neighborhood size. ERStream evaluates the pair-response tensor without materializing this Jacobian. The charge-response sum in Eq.~\eqref{eq:bec} contains two terms. The row term, $\sum_{j\neq i}\mathbf r_{ij}\otimes\partial\tq_i/\partial\mathbf r_{ij}$, collects the derivatives of the charge of atom $i$ with respect to the positions of its neighbors, which form row $i$ of the Jacobian $\partial\qraw/\partial\mathbf r$. The column term collects the derivatives of the neighboring charges with respect to the position of atom $i$, which form column $i$. The row term equals the derivative of $\tq_i$ with respect to a homogeneous strain $\boldsymbol\epsilon$ applied to the environment of atom $i$,
\begin{equation}
 \sum_{j\neq i}\mathbf r_{ij}\otimes\frac{\partial\tq_i}{\partial\mathbf r_{ij}}
 =\frac{\partial\tq_i}{\partial\boldsymbol\epsilon}\bigg|_{\boldsymbol\epsilon=0},\qquad
 \mathbf r_{ij}\to(\mathbf I+\boldsymbol\epsilon)\,\mathbf r_{ij},
 \label{eq:strain}
\end{equation}
where $\mathbf r_{ij}=\mathbf r_j-\mathbf r_i$ and $j$ runs over the dependency neighborhood of atom $i$. Because $\tq_i$ is rotation-invariant, this tensor is symmetric, and ERStream obtains its six independent components by propagating six strain tangents in forward mode (JVP) through the rebuilt tile.
ERStream accumulates the column term in a separate centered evaluation of the same tile. The atom-index symmetrization in Eq.~\eqref{eq:bec} is applied when the contributions of all tiles are summed.
The force contraction, column evaluation, and six tangent passes execute sequentially on one rebuilt tile, and each pass releases its computational graph before the next begins.
Each tangent pass costs about one forward evaluation of the tile, and the peak memory is bounded by a single computational graph of the tile. On several GPUs, the same tensor is obtained for any dependency reach from four tile VJPs and nine strain JVPs (SI Sec.~5).

For MACE-ER-T1 and DPA4C-ER, the dependency reach is one cutoff and the Jacobian is edge-local. Each learned charge depends only on the edge vectors $\mathbf r_{ij}$ of its own atom, and one further reverse-mode pass with a unit cotangent on the charge output returns every edge derivative $\partial\tq_i/\partial\mathbf r_{ij}$ of the tile. Both terms of Eq.~\eqref{eq:bec} follow from these edge derivatives. In MACE-ER-T1 the force and response passes share the rebuilt tile graph. The strictly local DPA4C-ER contracts the force and response terms in a single fused edge kernel (SI Sec.~2).

\method{Linear memory floor}
The persistent state of Eq.~\eqref{eq:floor} comprises the positions and velocities (48~bytes per atom), the raw charges and adjoints (16~bytes) and the forces, projected charges and pair-response tensors (104~bytes), together with $\kappa=K^{3}/N$ points of the FP64 SPME mesh per atom and $m_{\text{graph}}$ bytes per atom for the neighbor structure (Fig.~\ref{fig:method}b). MACE-ER holds a cell list of $O(N)$ storage as its only persistent neighbor state and builds the edges of a tile when the tile is evaluated. DPA4C-ER holds a compressed neighbor graph of 5~bytes per edge slot (SI Sec.~2). The tile workspace $W(B)$ of Eq.~\eqref{eq:nmax} follows from the halo factor of Eq.~\eqref{eq:halo}.

To approach $m_{\text{floor}}$ in practice, ERStream assigns explicit lifetimes to temporary system-sized buffers, stores the SPME reciprocal-space grid as a half-spectrum in a shared FFT workspace, restricts the intermediate buffers of the output heads to the owned atoms, and evaluates the memory model of Eq.~\eqref{eq:nmax} for every stage before allocation (SI Sec.~2 and 5).

\method{Free-space electrostatics}
Under open boundary conditions, periodic images and the reciprocal-space SPME mesh are absent, and only the global stage of Eq.~\eqref{eq:chain} changes. The long-range energy is the erf-regularized Coulomb sum
\begin{equation}
 E^{\text{LR}}=\frac{C}{2}\sum_{i\neq j}q_iq_j\,\frac{\operatorname{erf}(\alpha r_{ij})}{r_{ij}},
 \label{eq:freespace}
\end{equation}
where $C=1/(4\pi\varepsilon_0)=14.3996$~eV\,\AA\,$e^{-2}$ is the Coulomb constant, which also enters the periodic SPME energy, and $\alpha$ is the Gaussian screening parameter. The global stage evaluates Eq.~\eqref{eq:freespace} in FP64 together with the neutrality projection $\Pi_Q$ and the complete charge adjoint $\mathbf w$. The potential and its gradient at each site are computed in one pass per step and provide $E^{\text{LR}}$, $\mathbf w$, and $\mathbf F^{\text{LR}}$. The direct sum costs $O(N^{2})$ operations. For larger systems, a free-space solver splits the kernel as $\operatorname{erf}(\alpha r)/r=[\operatorname{erf}(\alpha r)-\operatorname{erf}(\beta r)]/r+\operatorname{erf}(\beta r)/r$ with $\beta<\alpha$. The first term is short-ranged and is summed over pairs within a cutoff. The second term is smooth and is evaluated on a uniform mesh by a zero-padded FFT convolution~\cite{hockney1988}. Atoms outside the mesh box interact with every atom through the direct sum. The split lowers the cost to $O(N\log N)$ at a bounded deviation from Eq.~\eqref{eq:freespace} (SI Sec.~5). Single-point evaluations that exceed device memory partition the direct sum through host memory.

\method{Multi-GPU execution}
The partition at the electrostatic barrier also defines how a step is distributed over several GPUs.  Each GPU runs one process of the same program (single program, multiple data), evaluates the local network for its share of the atoms together with their halos, and takes part in the exchange of per-atom vectors around the global solve. MACE-ER distributes tiles. Each GPU evaluates every $P$-th tile of a spatially ordered tile sequence, where periodic cells are divided into the boxes of a fractional-coordinate grid, reordered so that the tiles of each GPU carry similar cost, and isolated systems into Morton-ordered chunks of $B$ atoms. Halo atoms and layer features remain on the GPU that evaluates them, and no GPU holds the edge list of the whole system. All GPUs take part in the global solve, through a slab decomposition of the SPME reciprocal sum for periodic cells and of the long-range part of the split free-space solver for isolated systems. DPA4C-ER, which is strictly local, assigns one spatial domain with its cutoff halo to each GPU in periodic cells and evaluates the global solve on the first GPU. For both backbones, the remaining exchange of a step is confined to system-sized vectors. Positions and charge adjoints are broadcast from the first GPU over the NVIDIA Collective Communications Library (NCCL), and site energies, raw charges and forces of the owned atoms are reduced onto it. In FP64 the exchange amounts to
\begin{equation}
 m_{\text{x}}=\underbrace{24+8}_{\text{broadcast}}+\underbrace{16+24}_{\text{sum}}=72~\mathrm{bytes~per~atom}
 \label{eq:exchange}
\end{equation}
with a further 72~bytes per atom when the pair-response contributions are summed. A distributed solve adds charge broadcast and the reduction of partial potentials and fields (40~bytes per atom for SPME), plus transposes of the charge mesh, which scale with the number of mesh points. None of these terms depends on the feature dimension $d$ or the interaction depth $T$. Exchanging layer features at domain boundaries instead grows with both.

The step time on $P$ GPUs decomposes as
\begin{equation}
 t(P)\simeq t_{\mathrm G}+\frac{t_{\mathrm L}}{P}+t_{\mathrm x}(P),
 \label{eq:strong}
\end{equation}
where $t_{\mathrm G}$ is the part of the global solve that does not shrink with $P$, $t_{\mathrm L}$ the evaluation of all tiles on one GPU, and $t_{\mathrm x}$ the exchange of Eq.~\eqref{eq:exchange}. The tile workspace $W(B)$ and the halo factor $h(B)$ of Eq.~\eqref{eq:halo} apply per GPU. The first GPU additionally holds the system-sized state of Eq.~\eqref{eq:floor} and receives the reductions, so its memory sets the capacity on several GPUs. The tiled multi-GPU evaluation reproduces the whole-graph evaluation to FP64 round-off. The global solve bounds the speed-up by $1+t_{\mathrm L}/t_{\mathrm G}$.

\method{DFT calculations}
We generated candidate configurations for Interface MACE-ER with MD using foundation models, then selected 1,935 structures by applying DIRECT sampling separately to each structural category~\cite{batatia2025macemh1, wood2025uma, qi2024direct}. We labeled these configurations at the RPBE-D3(BJ) level~\cite{hammer1999rpbe, grimme2011d3bj}: GPU4PySCF for 220 isolated clusters~\cite{wu2025gpu4pyscf, li2025gpu4pyscf}, VASP for 385 solid configurations~\cite{kresse1996}, and ABACUS for 1,330 liquid, ice and interface configurations~\cite{li2016abacus}. 
The DFT labels for the 1,345 slab and interface configurations were computed with 20~\AA{} of vacuum and a dipole correction normal to the slab (SI Sec.~1). We trained Interface MACE-ER on energies and forces with a separate short-range energy head for each code and one charge head shared across all codes. Frames labeled by the same code use the same short-range head regardless of the long-range sum. 
During training, we evaluated the long-range energy with the direct erf-screened Coulomb sum for isolated clusters, with three-dimensional SPME for fully periodic frames, and with SPME and the Yeh--Berkowitz correction for slab frames~\cite{yeh1999slab}.
For the SrTiO$_3$ Born effective charges and electronic dielectric tensors, we performed PBE DFPT calculations with VASP for 30 structures, using the pseudopotentials and settings of the training labels~\cite{kresse1996, kresse1999}.

\section*{Data availability}
The water, SrTiO$_3$, C--H--O--N, BioMol, and qNEP water datasets are available from Refs.~\cite{schmiedmayer2024, zhong2025, wisesa2025, les2025, eastman2023spice, qnep2026} and the repositories cited there. The chignolin structure and NMR data are available from the Protein Data Bank (1UAO) and the Biological Magnetic Resonance Data Bank (5694). All datasets, trained models, and analysis code generated in this work will be released upon acceptance.

\section*{Code availability}
ERStream is available at \url{https://github.com/AM3GroupHub/ERStream} under the GNU Lesser General Public License v3.0 and will be archived on Zenodo upon publication. The MACE-ER and DPA4C-ER model code is available at \url{https://github.com/AM3GroupHub/mace-lr} and \url{https://github.com/AM3GroupHub/deepmd-lr}.

\section*{Acknowledgments} 
This work was supported in part by the AI2050 program at Schmidt Sciences (Grant G-25-69776 and Computing Fund) and the NUS High Performance Computing facility (NUSREC-HPC-00001).

\section*{Author contributions}
X.Y. and P.Z. conceived and designed the research. 
X.Y. developed the ERStream engine and codebase. 
Y.C. contributed to large-scale MD simulations and interpreted the results.
X.Y. and P.Z. wrote the manuscript with input from Y.C.
All authors contributed to discussions.

\bibliography{references}
\end{document}